\documentclass[10pt]{article}

\usepackage[OE]{express}
\usepackage{wrapfig}
\usepackage{amsmath}
\usepackage{braket}

\begin{document}

\title{Faster light with competing absorption and gain}

\author{Jon D. Swaim and Ryan T. Glasser\authormark{*}}
\address{Physics and Engineering Physics Department, Tulane University, New Orleans, LA 70118, USA}
\email{\authormark{*}rglasser@tulane.edu} 

\begin{abstract}
We experimentally investigate the propagation of optical pulses through a fast-light medium with competing absorption and gain.  The combination of strong  absorption and optical amplification in a potassium-based four-wave mixing process results in pulse peak advancements up to $88\%$ of the input pulse width, more than $35 \times$ that which is achievable without competing absorption.  We show that the enhancement occurs even when the total gain of the four-wave mixer is unity, thereby rendering the medium transparent.  By varying the pulse width, we observe a transition between fast and slow light, and show that fast light is optimized for large pulse widths.
\end{abstract}


\section{Introduction}

The ability to control optical information in pulses is an important aspect of many optical communcation schemes~\cite{Mok2005, Khurgin2005, Camacho2007, Boyd2009}, enhanced sensing protocols~\cite{Shi2007, Shahriar2007, Salit2007}, and experiments investigating fundamental physics~\cite{Stenner2003, Solli2004, Stenner2005}.  In general it is desirable to achieve tunable pulse delays or advancements upwards of the pulse width $\tau$ or more, while at the same time minimizing pulse distortion and reductions in the signal-to-noise ratio.  Dispersive effects in optical media with narrow-band resonances, such as gain lines~\cite{Wang2000, Dogariu2001, Glasser2012, Glasser2012a}, absorption lines~\cite{Chu1982, Camacho2007}, transparency features~\cite{Hau1999, Budker1999, Kash1999} and optical cavities~\cite{Tomita2011, Asano2016},  have been studied extensively for this purpose.  As an example, it has been shown in an ultracold atomic gas that light can be delayed by more than $5\tau$, corresponding to a group velocity of only $v_g = 17$ m/s~\cite{Hau1999}.  Additionally, the current record for slowest group velocity ($v_g = 8$ m/s in ~\cite{Budker1999}) was achieved with ms-length pulses in a rubidium-based electromagnetically induced transparency (EIT) experiment, and the fractional delay was 6\%.  On the other hand, experiments on superluminal pulse propagation (with $v_g > c$ or $v_g < 0$) have seen less progress in comparison.  While advancements on the order of $\tau$ have been reported using anomalous dispersion in absorbing media~\cite{Chu1982} and high-Q optical cavities~\cite{Asano2016}, these approaches are limited by the fact that the optimal advancement occurs at the point of minimum transmission, and as a result the advanced pulses are less than $1\%$ of the original pulse intensity.  One notable exception is a four-wave mixing (4WM) experiment which achieved advancements of 0.63 $\tau$ in the amplifying regime~\cite{Glasser2012a}.

It has been shown that modifications to the gain lineshape, e.g., asymmetric gain lines, can enhance the superluminal effects which arise due to anomalous dispersion on the sides of a gain resonance\cite{Glasser2012, Glasser2012a}.  Here, we show that the enhancement can occur in a nonlinear medium with regions of overlapping absorption and gain, without a modification to the lineshape.  We demonstrate this effect using 4WM in warm potassium vapor, which has a smaller ground state hyperfine splitting than rubidium or cesium and generally leads to strong absorption of the probe pulse, even at room temperature~\cite{Zlatkovic2016, Swaim2017, Swaim2017a}.  Because the probe pulse is effectively positioned in the center of the absorption line, rather than on the wing \cite{Glasser2012, Glasser2012a}, the shapes of the 4WM resonances remain symmetric in frequency.  Using this approach, we observe relative advancements up to ($0.88 \pm 0.04$) $\tau$, and show that the enhancement occurs even when the medium has become transparent (i.e., characterized by an effective gain $g_{\textrm{eff}} = 1.0 \pm 0.2$).  We note that the strong superluminal effects reported here have not been previously observed in the transparent regime, and that they also occur for the generated conjugate pulse, despite the fact that the absorption features are not present at the off-resonance conjugate frequency.   

\section{Experimental setup}

\begin{figure}[h!]
\centering
\includegraphics[width=\textwidth]{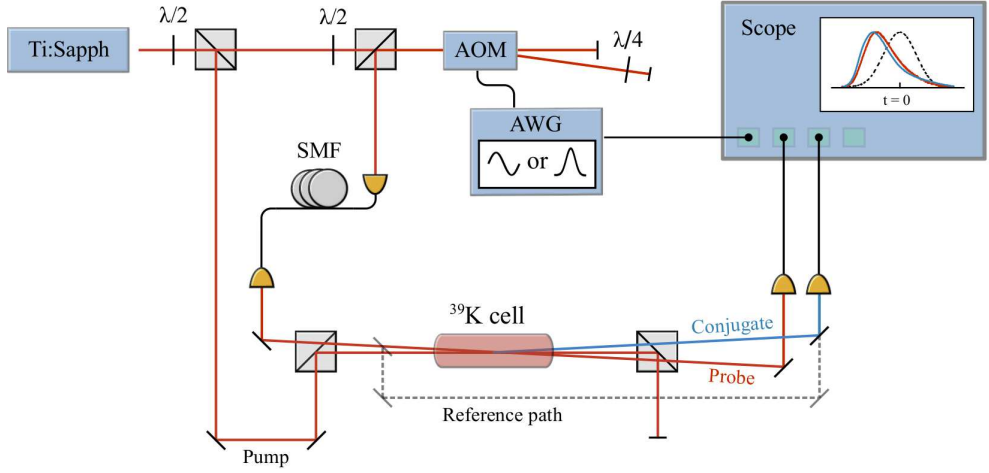}
\caption{\label{setup}  Setup for generating fast light.  Four-wave mixing occurs when a strong pump field interacts with a weak, detuned probe in the center of a $^{39}$K atomic vapor cell.  The input probe pulses are produced by passing some of the light through an AOM which is pulsed with an AWG.  The 4WM conjugate and probe pulses are then detected and compared with a reference pulse which is directed around the vapor cell.  $\lambda/2$: half-wave plate; $\lambda/4$: quarter-wave plate; AOM: acoustic-optical modulator; AWG: arbitrary waveform generator; SMF: single mode fiber.}
\end{figure}

The experimental setup for generating pulsed 4WM in warm potassium vapor is based on previous 4WM experiments~\cite{McCormick2007, Glasser2012, Zlatkovic2016, Swaim2017, Swaim2017a} and shown schematically in Fig.~\ref{setup}.  Briefly, a strong, linearly polarized pump field ($P \sim 380$ mW, $\lambda \sim 770$ nm) is detuned approximately $900$ MHz to the blue side of the $^{39}$K D1 line and sent through an 80 mm, anti-reflection-coated potassium vapor cell, kept at a temperature of $T \approx 110 ^\circ$C.  The pump is slightly elliptical (spatially) with a size of $0.8$ mm $\times$ $0.6$ mm, and overlaps with a weak probe field ($1/e^2$ diameter $\approx 670$ $\mu$m) in the center of the cell at an angle of $\sim 0.1 ^\circ$.  As the effective interaction strength of the process increases with pump power and atomic density (temperature) these parameters were optimized for pulse advancement and minimal distortion.  The probe is generated by downshifting a fraction of the pump light with an acousto-optic modulator (AOM) operating near $\Omega/2\pi = 220$ MHz in the double-pass configuration, resulting in a probe that is detuned $\sim 440$\,MHz blue of the pump, and then pulsing the AOM using an arbitrary waveform generator.  The probe is also linearly polarized, orthogonal to the pump beam.  After passing through a single-mode fiber to clean the mode, the average power of the probe pulse is $P_{\textrm{in}} \sim 10$ $\mu$W.   In addition to amplifying the probe pulse, the 4WM process generates a second, conjugate pulse, detuned from the pump by $2\Omega$, which exits the cell at an angle of $\sim 0.2 ^\circ$ relative to the probe.  The pump is filtered out using a Glan-Taylor polarizer, and the probe and conjugate pulses are sent to photodetectors, amplified and averaged $128$ times on an oscilloscope.  In each experiment, a reference pulse traveling at the speed of light in air is obtained by redirecting the probe around the cell using a flip-mirror.

\section{Results}

\begin{figure}[h!]
\centering
\includegraphics[width=\textwidth]{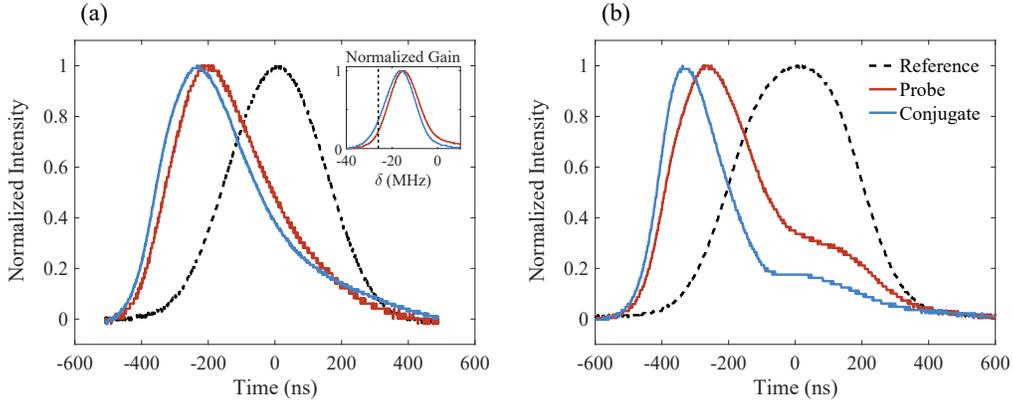}
\caption{\label{fastlight} Fast light.  (a) The probe and conjugate are advanced by $76\%$ and $88\%$ with respect to the input pulse of width $\tau = 260$ ns.  The 4WM resonances are shown in the inset, and the dashed vertical line indicates the two-photon detuning where fast light occurs.  (b)  Advancements in the transparent regime, accompanied by some additional pulse distortion.}
\end{figure}

Fast light occurs when the probe frequency is tuned to the side of the gain resonance, as shown in the inset of Fig.~\ref{fastlight}(a).  We show two sets of superluminal pulses in Fig.~\ref{fastlight} for this case, where we have used $\delta = -26$ MHz and $\delta = -24$ MHz in (a) and (b), respectively.  In the first result, the peaks of the probe and conjugate pulses are advanced by $0.76 \tau$ and $0.88 \tau$, respectively, based on an input pulse width of $\tau = 260$ ns.  These advancements are greater than those reported previously in a four-wave mixing scenario~\cite{Glasser2012a}, and further differ in that both the gain resonances are symmetric in frequency (the gain resonances observed here are Gaussian in shape, not Lorentzian).  Given that the length of the vapor cell is $L = 80$ mm, these advancements correspond to group velocities of $v_g = -1.3 \times 10^{-3} c$ and $v_g = -1.2 \times 10^{-3} c$, respectively.  Also, we note that the result is obtained when the effective gain of the four-wave mixer (sum of the output intensities divided by the input) is only $g_{\textrm{eff}} = 1.8 \pm 0.2$.  

In the result shown in Fig.~\ref{fastlight}(b), we increase the average pulse power to $P_{\textrm{in}} \sim 17$ $\mu$W and use $\tau = 390$ ns, which has the effect of increasing the absolute advancement as well as the pulse distortion, though the fractional advancements are slightly less in this case ($0.67 \tau$ and $0.84 \tau$).  Here, we are able to considerably advance the pulses even when the medium has become transparent due to competing absorption and gain ($g_{\textrm{eff}} = 1.0 \pm 0.2$), which to our knowledge is the first demonstration of such an effect.  In this manner, the fast-light medium may be thought of as a combined beamsplitter and frequency shifter, resulting in the fast-light propagation of both the probe pulse and frequency-shifted conjugate pulse, with total output intensity being equal to the input.  Although transparent gain-assisted superluminal light was demonstrated in ~\cite{Dogariu2001}, in that experiment the fractional advancement was only several percent (and involved a different, double-gain line system and no conjugate generation).  In this experiment we aimed to adjust only the 4WM parameters to achieve the desired result. That is, there is no artificially applying separate attenuation to achieve transparency. Ultimately, then, there were several choices of parameters which gave unity gain transmission (frequencies of the pump and probe, for example), with varying degrees of advancement and distortion.  

\begin{figure}[t!]
\centering
\includegraphics[width=\textwidth]{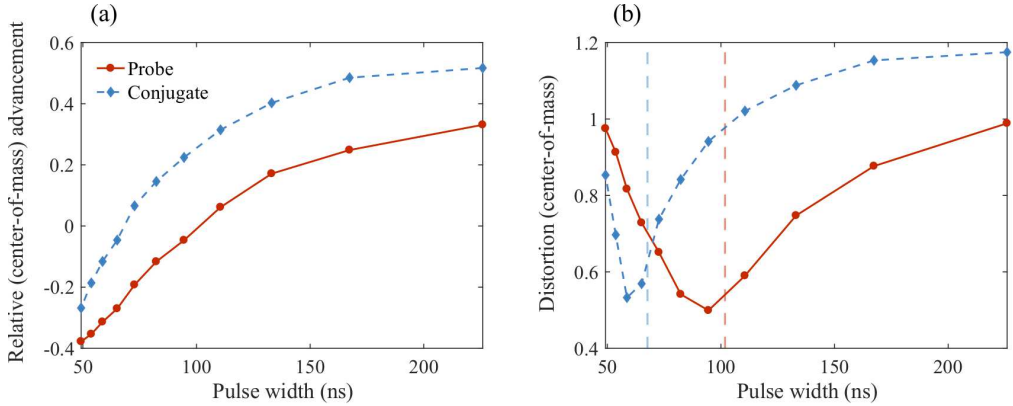}
\caption{\label{bandwidths} Tunable advancements and pulse bandwidth.  (a) When 4WM is tuned for fast light, superluminality is optimized for longer pulse widths.  (b)  Pulse distortion associated with both fast and slow light is minimized for bandwidths close to those which give zero advancement (indicated with dashed, vertical lines).  }
\end{figure}

A simple method of tuning the advancement is found by varying the pulse bandwidth, or $\tau$.  To demonstrate this, we vary $\tau$ from $50$ ns to $225$ ns and measure the output pulses, keeping the peak input pulse intensity and 4WM parameters fixed.  Since, in general, the pulses can be distorted, we calculate the advancements in terms of the center-of-mass of the pulse and show the result in Fig.~\ref{bandwidths}(a).  As expected, fast light is optimized for larger pulse widths on account of the smaller bandwidths and uniform dispersion.  For shorter $\tau$, the pulses have frequency components which coincide with the peaks of the gain resonances, and slow light occurs.  Thus, we observe a transition between slow and fast light, which differs for the probe and conjugate pulses since they are coupled in the 4WM process (the gain resonances are slightly shifted in frequency).  This result demonstrates the tunability of the dispersion in 4WM.  In addition, we calculate the pulse distortion for these experiments, using the formula given in ~\cite{Macke2003}.  As shown in Fig.~\ref{bandwidths}(b), increased distortion accompanies both advancement and delay, with the minimum occuring for a pulse width which is close to that which gives zero advancement, as indicated by dashed vertical lines.  The fact that the distortion increases on either side, for both slow and fast light, is in agreement with the view that the mechanism for both advancement and delay is a pulse-reshaping (that is, a re-phasing) phenomenon.  We note that even in the case of pulse advancement, the speed of information transfer is limited to the speed of light in vacuum \cite{Swaim2017a}.

\section{Discussion}

A distinguishing characteristic of potassium is the relatively small hyperfine ground state splitting ($\Delta_{\textrm{HFS}} = 462$ MHz), which naturally leads to strong absorption of the probe.  In combination with the large $\chi^{(3)}$ nonlinearity in potassium, there then exists a strong competition between absorption and gain for a variety of experimental parameters.  We found that this competing absorption serves to enhance superluminality.   To illustrate this, we can calculate the expected advancement from a gain resonance in the absence of absorption.   We consider a simplified, toy model where the pulse propagates according to its frequency-dependent wavevector

\begin{equation}
k(\omega) = \left(\omega/c\right) n(\omega) = \left(\omega/c\right) - \frac{\textrm{ln}(g_0)/L \gamma_0}{2\left(\omega - \omega_0 +i\gamma_0\right)}
\end{equation}

\noindent where $c$ is the speed of light in vacuum, and the resonant features of the medium's refractive index $n(\omega)$ are given by a Lorentzian with peak gain $g_0$, linewidth $\gamma_0$ and frequency $\omega_0$.  In this model the pulse advancement is given by $\Delta t = L \left(\textrm{Re}[dk/d\omega] \right)-L/c$.  Taking our best fast light result in Fig.~\ref{fastlight}(a) as an example (with $g_0 = 5$, $\omega_0 = -17$ MHz and $\gamma_0 = 16$ MHz), we find a \emph{maximum} predicted advancement of only $6.3$ ns using this gain-only model.  Thus, our results correspond to an enhancement of more than $35\times$ when compared to that which can be achieved using only gain. As fast-light is expected near the center of an absorption line, as well as near the wings of a gain line \cite{Glasser2012}, the current scheme demonstrates a significant enhancement of superluminality when such features are overlapped and reinforce one another.  

It is worthwhile to mention that the approach taken here is compatible with a gain doublet configuration, which can act to reduce pulse distortion provided that the gain is spectrally flat near the doublet center~\cite{Steinberg1994, Wang2000, Dogariu2001}.  Such an approach could provide a further improvement to the results shown here, and could be significant given that our advancements are naturally associated with distortion.  Still, we find that our superluminal pulses can be less distorted than those demonstrated in a motivating experiment in rubidium~\cite{Glasser2012a}, which could in part be ascribed to symmetric gain resonances.  It is also worth mentioning that input pulse power~\cite{Glasser2012} and spatial dispersion~\cite{Torres2010, Glasser2012a, Swaim2017a} contribute to both the advancement and distortion.  So, future experiments might aim to quantify these effects and attempt to find the best methods of distortion management. 

Lastly, we would like to highlight the fact that our four-wave mixer generates superluminal light at a new frequency (i.e., the conjugate frequency), even though, overall, the medium is effectively transparent.  This contrasts with an earlier demonstration of fast light through a transparent medium, in which an input pulse traveling at $c$ was made superluminal~\cite{Dogariu2001}.  In fact, we found it to be generally true that the largest advancements occurred for the conjugate pulse.  This is consistent with the experiments in ~\cite{Glasser2012, Glasser2012a}, where the largest advancement also occurred for the blue detuned pulse (though in those experiments it was referred to as the probe pulse) and is likely a result of the coupling between probe and conjugate fields in 4WM.
  
\section{Conclusion}

We have demonstrated that competing absorption and gain in a nonlinear medium can lead to the transmission and generation of optical pulses with enhanced superluminality, even in the regime of unity gain. With optimization based on pulse width and frequency detunings, we achieve a maximum relative advancement of $0.88\tau$ for a pulse width of $\tau = 260$ ns.  In this new regime, the medium's dispersion can be tailored to outperform equivalent setups which rely solely on either absorption or gain, where in particular we calculate an improvement of more than $35\times$ over the gain-only configuration.    Overall, these results suggest that with such an approach it may be possible to achieve even greater control over optical pulses, without introducing adverse effects such as additional pulse distortion and attenuation. 

\section*{Funding}
Louisiana State Board of Regents (Grant 073A-15); Northrop Grumman -- NG NEXT.

\section*{Acknowledgments}
The authors acknowledge useful discussions with Erin M. Knutson and Wenlei Zhang.

\end{document}